\documentclass[12pt,epsf,amssymb,qsymbols]{article}
\usepackage{tabularx}
\usepackage{array}
\usepackage{graphics}
\usepackage{graphicx}
\usepackage{epsfig}
\usepackage{amsmath}
\usepackage{amssymb}
\usepackage{psfrag} 
\usepackage{color}
\makeatletter

\usepackage{verbatim}

\newcommand{\nn}{\nonumber}
\newcommand{\bea}{\begin{eqnarray}}
\newcommand{\eea}{\end{eqnarray}}
\newcommand{\beq}{\begin{equation}}
\newcommand{\eeq}{\end{equation}}

\newcommand{\gev}{{\rm GeV}}

\newcommand{\pdir}{p\kern -5.2pt\raise 0.2ex\hbox {/}}
\newcommand{\vdir}{v\kern -5.75pt\raise 0.15ex\hbox {/}}
\newcommand{\kdir}{k\kern -5.75pt\raise 0.15ex\hbox {/}}
\newcommand{\epsdir}{\epsilon\kern -5.0pt\raise 0.15ex\hbox {/}}
\newcommand{\bvdir}{\bar{v}\kern -5.75pt\raise 0.15ex\hbox {/}}
\newcommand{\Ddir}{D\kern -7.75pt\raise 0.20ex\hbox {/}}
\newcommand{\ldir}{l\kern -5.0pt\raise 0.2ex\hbox{/}}
\newcommand{\varepsdir}{\varepsilon\kern -5.5pt\raise 0.15ex\hbox{/}}

\newcommand{\bbarq}{B^0_q-\overline B^0_q}
\newcommand{\bbars}{B^0_s-\overline B^0_s}
\newcommand{\bbard}{B^0_d-\overline B^0_d}

\def\negcdot{\negmedspace\cdot\negmedspace}
\makeatother

\begin{document}
\thispagestyle{empty}
\begin{flushright}
\begin{tabular}{l}
\scalebox{.83}{\tt LPT Orsay, 06-94}
\end{tabular}
\end{flushright}
\begin{center}
\vskip 3.7cm\par
{\par\centering \Large \bf Chiral behavior of the $B^0_{d,s}-\overline B^0_{d,s}$ mixing}\\
\vskip 0.18cm\par
{\par\centering \Large \bf amplitude in the Standard Model and beyond}\\
\vskip 1.75cm\par
\scalebox{.89}{\par\centering \large  
\sc Damir~Be\'cirevi\'c$^a$, Svjetlana~Fajfer$^{b,c}$ and Jernej~Kamenik$^{b}$}
{\par\centering \vskip 0.5 cm\par}
{\sl
$^a$ Laboratoire de Physique Th\'eorique (B\^at 210)~\footnote{Unit\'e mixte de
Recherche du CNRS - UMR 8627.}, Universit\'e Paris Sud, \\
Centre d'Orsay, 91405 Orsay-Cedex, France.\\                                   
\vspace{.25cm}
$^b$ J.~Stefan Institute, Jamova 39, P.O. Box 3000,\\
1001 Ljubljana, Slovenia.\\
%\vskip1.cm
\vspace{.25cm}
$^c$
Department of Physics, University of Ljubljana,\\
 Jadranska 19, 1000
Ljubljana,
Slovenia.  }\\
{\vskip 0.25cm \par}
\end{center}

\vskip 1.75cm
\begin{abstract}
We compute the chiral logarithmic corrections to the $\bbard$ and $\bbars$ mixing amplitudes 
in the Standard Model and beyond. We then investigate the impact of the inclusion of  
the lowest-lying scalar heavy-light states to the decay constants and bag-parameters 
and show that this {\bf does not} modify the pion chiral logarithms, but it does produce corrections 
which are competitive in size with the $K$- and $\eta$-meson chiral logarithms. 
This conclusion is highly relevant to the lattice studies since the pion
 chiral logarithms represent the most important effect in guiding the chiral 
 extrapolations of the lattice data for these quantities.  It is also important to 
 stress that the pion chiral logarithmic corrections are useful in guiding those chiral 
 extrapolations as long as $m_\pi \ll \Delta_S$, where $\Delta_S$ stands for the mass difference between the heavy-light mesons belonging to $\frac{1}{2}^+$ and $\frac{1}{2}^-$ doublets.
\end{abstract}
\vskip 0.2cm
\setcounter{page}{0}
\setcounter{footnote}{0}
\setcounter{equation}{0}
%%%%%%%%%%%%%%%%%%%%%%%%%%%%%%%%%%%%%%%%
%%%%%%%%%%%%%%%%%%%%%%%%%%%%%%%%%%%%%%%%
%%%%%%%%%%%%%%%%%%%%%%%%%%%%%%%%%%%%%%%%
\noindent

\renewcommand{\thefootnote}{\arabic{footnote}}
%\vspace*{-1.5cm}

\newpage
\setcounter{footnote}{0}
%%%%%%%%%%%  Section 1

\section{Introduction\label{sec:0}}

The oscillations in the $B^0_{d,s}-\overline  B^0_{d,s}$ systems  are mediated by the flavor changing neutral currents which  are forbidden at tree level of the Standard Model (SM) and therefore their detection gives access to the particle content in the corresponding loop diagrams.  First experimental measurement of a {\it large} value of $\Delta m_{B_d}$ indicated that the top quark mass was very heavy~\cite{Albajar:1986it}, which was confirmed almost a decade later in the direct measurements, $m_t=
172.5\pm 1.3\pm 1.9 \ \gev$ through the $p\bar p$-collisions~\cite{Mtop}. 
Nowadays, the accurately measured  $\Delta m_{B_d}=0.509(5)(3)\ {\rm ps}^{-1}$~\cite{hfag}, and $\Delta m_{B_s}=17.31(^{33}_{17})(7)\ {\rm ps}^{-1}$~\cite{cdf}, are used to constrain the shape of the Cabbibo-Kobayashi-Maskawa (CKM) unitarity triangle and thereby determine the amount 
of the CP-violation in the SM~\cite{UTA}.  This goal is somewhat hampered  by the theoretical 
uncertainties in computing the values for the two decay constants, 
$f_{B_{s,d}}$, and the corresponding ``bag" parameters, $B_{B_{s,d}}$. These quantities can, in principle, be  computed on the lattice.~\footnote{Recent reviews on the current status of the lattice QCD computations of $\bbarq$ mixing amplitudes can be found in ref.~\cite{onogi}.} However, a major obstacle in the current lattice studies is that the $d$-quark cannot be reached directly but through an extrapolation of the results obtained by working with larger light quark masses down to the physical $d$-quark mass. 
This extrapolation induces systematic uncertainties which are hard to control as the spontaneous chiral symmetry breaking effects are expected to become increasingly pronounced as one lowers the light quark mass~\cite{ryan}.   Heavy meson chiral perturbation theory (HMChPT) allows us to gain some control over these uncertainties because it predicts the chiral behavior of the hadronic quantities relevant to the heavy-light quark phenomenology which then can be implemented to guide the extrapolation of the lattice results.  
HMChPT combines the heavy quark effective theory  (HQET) with the common pattern of spontaneous breaking of the chiral symmetry, $SU(3)_L\otimes SU(3)_R$ $\to$
$SU(3)_V$~\cite{casalbuoni}. 

Like in the standard ChPT, in HMChPT one computes the chiral logarithmic corrections (the so-called non-analytic terms) which are expected to be relevant to the very low energy region, i.e.,  
$m_q \ll \Lambda_{\rm QCD}$. 
While this condition is satisfied for  $u$- and $d$-quarks, the situation with the $s$-quark is still unclear~\cite{seb}.  Also ambiguous is the size of the chiral symmetry breaking scale, $\Lambda_\chi$. Some authors consider it to be around $4\pi f_\pi \simeq 1$~GeV~\cite{manohar}, while others prefer identifying it with the mass of the 
first vector resonance, $m_\rho=0.77$~GeV~(see e.g. ref.~\cite{pich}), and sometimes even lower~\cite{gasser}. In the heavy-light quark systems the situation becomes more complicated because the first orbital excitations  ($j_\ell^P=1/2^+$) are not far away from the lowest lying states ($j_\ell^P=1/2^-$).  
The recent experimental evidence for the scalar $D_{0s}^{\ast}$ and axial $D_{1s}$ mesons indicate that this splitting is only $\Delta_{S_s} \equiv m_{D_{0s}^{\ast}}-m_{D_s}=m_{D_{1s}}-m_{D_s^\ast}=350$~MeV~\cite{Ds}, and somewhat larger for the non-strange states $\Delta_{S_{u,d}}=430(30)$~MeV~\cite{Dq}.~\footnote{We note, in passing, that the experimentally established fact,  $\Delta_{S_s} < \Delta_{S_{u,d}}$, 
is not yet understood~\cite{dss} although a recent lattice study with the domain wall quarks indicates a qualitative agreement with experiment~\cite{norikazu}.  }  This and the result of the lattice QCD study in the static heavy quark limit~\cite{green} suggest that the size of this mass difference remains as such in the $b$-quark sector as well. One immediately observes that both  $\Delta_{S_s}$ and  $\Delta_{S_{u,d}}$ are  smaller 
than $\Lambda_\chi$, $m_\eta$, and even $m_K$, which requires revisiting the predictions based on HMChPT and reassessing their range of validity.  In this paper we investigate this 
issue on the specific examples of the decay constants $f_{B_{d,s}}$ and the
bag parameters which enter the investigation of the SM and supersymmetric (SUSY) 
effects in the $\bbard$ and $\bbars$ mixing amplitudes~\cite{SUSY-B}.

\section{Bases of $\Delta B=2$ operators and $B$-parameters\label{sec:1}}

The SUSY contributions to the $\bbarq$ mixing amplitude, where $q$ stands for either $d$- or $s$-quark, 
are usually discussed in the so called SUSY basis of $\Delta B=2$ operators~\cite{gabbiani}:
\bea
 \label{baseS}
{\phantom{{l}}}\raisebox{-.16cm}{\phantom{{j}}}
O_1 &=& \ \bar b^i \gamma_\mu (1- \gamma_{5} )  q^i \,
 \bar b^j  \gamma^\mu (1- \gamma_{5} ) q^j \,  , 
  \nonumber \\
{\phantom{{l}}}\raisebox{-.16cm}{\phantom{{j}}}
O_2&=& \ \bar b^i  (1- \gamma_{5} ) q^i \,
\bar b^j  (1 - \gamma_{5} )  q^j \, ,  \nonumber  \\
{\phantom{{l}}}\raisebox{-.16cm}{\phantom{{j}}}
O_3&=& \ \bar b^i  (1- \gamma_{5} ) q^j \,
 \bar b^j (1 -  \gamma_{5} ) q^i \, ,  \\
{\phantom{{l}}}\raisebox{-.16cm}{\phantom{{j}}}
O_4 &=& \ \bar b^i  (1- \gamma_{5} ) q^i \,
 \bar b^j   (1+ \gamma_{5} ) q^j \,  ,  \nonumber \\
{\phantom{{l}}}\raisebox{-.16cm}{\phantom{{j}}}
O_5 &=& \ \bar b^i  (1- \gamma_{5} ) q^j \,
 \bar b^j   (1+ \gamma_{5} ) q^i \,  ,  \nonumber 
 \eea
where $i$ and $j$ are the color indices.  Although the operators in the above bases are written with both parity even and parity odd parts, only the 
parity even ones survive in the matrix elements. In SM,   only  $O_1$ (left-left) operator is relevant in 
describing  the $\bbarq$ mixing amplitude. 
The matrix elements of the above operators are conventionally parameterized 
in terms of bag-parameters, $B_{{1}{\mathrm -}{5}}$, as a measure of the discrepancy with respect to 
the estimate obtained by using the vacuum saturation approximation (VSA), 
\bea
{\langle \bar B^0_q\vert O_{{1}{\mathrm -}{5}}(\nu)\vert B^0_q\rangle 
\over 
\langle \bar B^0_q\vert O_{{1}{\mathrm -}{5}}(\nu)\vert B^0_q\rangle_{\rm VSA} } = B_{{1}{\mathrm -}{5}}(\nu)\,,
\eea 
where $\nu$ is the renormalisation scale of the logarithmically divergent operators, $O_i$,
 at which the separation between the long-distance (matrix elements) and short-distance 
 (Wilson coefficients) physics is made.  We remind the reader that
\bea
 \label{params}
\langle \bar B^0_q \vert  O_1 \vert  B^0_q \rangle_{\rm VSA}    &=& 2 \left( 1 + {1\over 3}\right)  \langle \bar B^0_q \vert A_\mu\vert 0\rangle\   \langle  0 \vert A^\mu \vert B^0_q \rangle \,,  \nonumber \\
\langle \bar B^0_q \vert  O_2 \vert  B^0_q \rangle_{\rm VSA}  &=& -2 \left( 1 - \frac{1}{6} \right)  \, \left|
 \langle 0 \vert P \vert B^0_q \rangle \right|^2\,,\nonumber  \\
\langle \bar B^0_q \vert  O_3 \vert  B^0_q \rangle_{\rm VSA} &=&  \left( 1 - \frac{2}{3} \right)  \, \left|
 \langle 0 \vert P \vert B^0_q \rangle  \right|^2\,,  \\
\langle \bar B^0_q \vert  O_4 \vert  B^0_q \rangle_{\rm VSA}  &=&  {1\over 3}  \langle \bar B^0_q \vert A_\mu\vert 0\rangle\   \langle 0 \vert A^\mu\vert B^0_q \rangle + 2 \left|
 \langle 0 \vert P \vert B^0_q \rangle \right|^2\,,\nonumber  \\
\langle \bar B^0_q \vert  O_5 \vert  B^0_q \rangle_{\rm VSA}  &=&    \langle \bar B^0_q \vert A_\mu\vert 0\rangle\   \langle 0 \vert A^\mu\vert B^0_q \rangle + {2\over 3} \left|
 \langle 0 \vert P \vert B^0_q \rangle  \right|^2\,,\nonumber 
 \eea
with  $A_\mu = \bar b \gamma_\mu\gamma_5 q$ and $P = \bar b \gamma_5 q$ being the axial current and the pseudoscalar density, respectively. In HQET, in which we will be working from now on, the field $ \bar b$ is replaced by the static one, $h^\dagger$, which satisfies  $h^\dagger \gamma_0=h^\dagger$. This equation and the fact that the amplitude is invariant under the Fierz transformation in Dirac indices, eliminate the operator $O_3$ from further discussion, i.e., 
$\langle  \bar B^0_q\vert  \widetilde O_3 + \widetilde O_2 +\frac{1}{2} \widetilde O_1 \vert   B^0_q \rangle = 0$, where the tilde is used to stress that the operators are now being considered in the static limit of HQET ($\vert \vec v\vert =0$). Furthermore, in the same limit
\bea
\lim_{m_b\to \infty}{  \langle 0 \vert A_\mu \vert B^0_q(p) \rangle_{\rm QCD} \over \sqrt{2m_B}} =\lim_{m_b\to \infty}{ \langle 0 \vert P \vert B^0_q(p) \rangle_{\rm QCD} \over \sqrt{2m_B}}  =    \langle 0 \vert \widetilde A_0 \vert B^0_q(v) \rangle_{\rm HQET} = i \hat f_q \,,
\eea
where $\hat f_q$ is the decay constant of the static $1/2^-$ heavy-light meson, and the HQET states are normalized as $\langle B_q^0(v)\vert B_q^0(v^\prime)\rangle=\delta(v-v^\prime)$, 
so that we finally have
\bea
\label{hqet-basis}
\langle \bar B^0_q \vert  \widetilde O_1(\nu) \vert  B^0_q \rangle   &=&  {8\over 3} \hat f_q(\nu)^2  \widetilde B_{1q}(\nu)\,,  \nonumber \\
\langle \bar B^0_q \vert  \widetilde O_2(\nu) \vert  B^0_q \rangle   &=&  -{5\over 3} \hat f_q(\nu)^2  \widetilde B_{2q}(\nu)\,,  \, \\
\langle \bar B^0_q \vert  \widetilde O_4(\nu) \vert  B^0_q \rangle &=& {7\over 3} \hat f_q(\nu)^2  \widetilde B_{4q}(\nu)\,,\nonumber  \\
\langle \bar B^0_q \vert \widetilde  O_5(\nu) \vert  B^0_q \rangle &=&    {5\over 3} \hat f_q(\nu)^2  \widetilde B_{5q}(\nu)\,.\nonumber 
 \eea
One of the reasons why lattice QCD is the best currently available method for computing these matrix elements is the fact that it enables a control over the $\nu$-dependence by verifying the corresponding 
renormalisation group equations, which is essential for the cancellation against the $\nu$-dependence in the corresponding perturbatively computed Wilson coefficients~\cite{4f}.  From now on we will assume that the UV divergences are being taken care of and the scale $\nu$ will be implicit.

\section{Chiral logarithmic corrections}

In this section we use HMChPT to describe the low energy behavior of the matrix elements~(\ref{hqet-basis}). 
Before entering the details, we notice that the operators $\widetilde O_4$ and $\widetilde O_5$ differ only in the color indices, i.e., by a gluon exchange, which is a local effect that cannot influence the long distance behavior described by ChPT. In other words, from the point of view of ChPT, the entire difference of the chiral behavior of the bag parameters $\widetilde B_{4q}$ and $\widetilde B_{5q}$ is 
encoded in the local counter-terms, whereas their chiral logarithmic behavior is the same.  Similar 
observation has been made for the operators entering the SUSY analysis of the $\bar K^0$-$K^0$ mixing amplitude, as well as for the electromagnetic penguin operators in $K\to \pi\pi$ decay~\cite{giovanni}. 
Thus, in the static heavy quark limit ($m_Q\to \infty$),  we are left with the first three operators in eq.~(\ref{hqet-basis}) which, in their bosonised version, can be  written as~\cite{detmold-lin}~\footnote{In the first version of the present paper only the factorisable pieces in the bosonised forms of the operators $\widetilde  O_{2,4}$ were considered. We thank David Lin for pointing out to us the presence of the corresponding nonfactorisable pieces, recently considered in ref.~\cite{detmold-lin}, which are  being properly accounted for in this version of our paper.}
\bea\label{bosbase}
\widetilde  O_1 &=& \sum_X \beta_{1X} {\rm Tr} \left[ (\xi \overline H^Q)_q \gamma_{\mu}
(1-\gamma_5) X \right] {\rm Tr} \left[ (\xi H^{\bar Q})_q \gamma^{\mu} (1-\gamma_5) X \right] + {\rm c.t.}\,,\nn \\
 &&\hfill \nn \\
\widetilde O_2 &=&\sum_X \beta_{2X} {\rm Tr} \left[ (\xi \overline H^Q)_q (1-\gamma_5) X \right] {\rm Tr} \left[
(\xi H^{\bar Q})_q (1-\gamma_5) X \right]+ {\rm c.t.}\,,\nn \\
 &&\hfill \nn \\
\widetilde O_4 &=& \sum_X \beta_{4X} {\rm Tr} \left[ (\xi  \overline H^Q)_q (1-\gamma_5) X \right] {\rm Tr} \left[
(\xi^{\dagger} H^{\bar Q})_q (1+\gamma_5) X \right]  \nn \\
&&\hspace*{4mm} + \bar \beta_{4X} {\rm Tr} \left[ (\xi H^{\bar Q})_q
(1-\gamma_5) X \right]  {\rm Tr} \left[ (\xi^{\dagger}  \overline H^Q)_q (1+\gamma_5) X \right] + {\rm c.t.}\,,
\eea
where $X\in\{ 1, \gamma_5, \gamma_{\nu},
\gamma_{\nu}\gamma_5, \sigma_{\nu\rho}\}$.~\footnote{Contraction of Lorentz
indices and HQET parity conservation requires the same $X$ to appear in
both traces of a summation term. Any insertions of $\vdir $ can be
absorbed via $\vdir  H = H$, while any nonfactorisable
contribution with a single trace over Dirac matrices can be reduced to
this form by using the $4\times 4$ matrix identity 
\bea 4 {\rm Tr}(AB) &=& 
{\rm Tr}(A){\rm Tr}(B) + 
{\rm Tr}(\gamma_5 A){\rm Tr}(\gamma_5 B) +
{\rm Tr}(A\gamma_{\mu}){\rm Tr}(\gamma^{\mu}B) \nn \\
&&+{\rm Tr}(A\gamma_{\mu}\gamma_5){\rm Tr}(\gamma_5\gamma^{\mu}B) + 
1/2{\rm Tr}(A\sigma_{\mu\nu}){\rm Tr}(\sigma^{\mu\nu}B).\nn \eea} As before 
the index ``$q$" denotes the light quark flavor, and ``c.t." stands for the local counter-terms. To relate  $\beta_i$'s to the bag parameters in eq.~(\ref{hqet-basis}) we should recall that  the field $H_q$ is built
 up from the pseudoscalar ($P$) and the vector ($P^\ast$) meson fields as 
\bea\label{fieldsH}
H_q^{Q}(v) &=& {1+\vdir \over 2}\left[ P_\mu^{Q \ast }(v)\gamma^\mu - P^Q(v) \gamma_5\right]_q\,,\cr
H_q^{\bar Q}(v)& =& \left[ P_\mu^{\bar Q \ast}(v)\gamma^\mu - P^{\bar Q}(v) \gamma_5\right]_q{1-\vdir \over 2}\,,
\eea
thus exhibiting the heavy quark spin symmetry for the lowest lying $j_\ell^P=1/2^-$ states. After 
evaluating the traces in eq.~(\ref{bosbase}) we obtain 
\bea
\widetilde B_1 ={3\over 2\hat f^2}\widehat \beta_1\,,\quad 
\widetilde B_2 ={12\over 5\hat f^2}\widehat\beta_2\,,\quad 
\widetilde B_4 ={12\over 7\hat f^2}\widehat\beta_4\,,\quad 
\widetilde B_5 ={12\over 5\hat f^2}\widehat\beta_4\, ,
\eea 
where 
\bea\label{betas}
\widehat \beta_1 &=& \beta_1 + \beta_{1\gamma_5} - 4 (\beta_{1\gamma_{\nu}} +
\beta_{1\gamma_{\nu}\gamma_5}) - 12 \beta_{1\sigma_{\nu\rho}},\nn\\
\widehat \beta_2&=&-\beta_2 - \beta_{2\gamma_5} + \beta_{2\gamma_{\nu}} +
\beta_{2\gamma_{\nu}\gamma_5},\nn\\
\widehat \beta_4 &=& \beta_4 - \beta_{4\gamma_5} - \beta_{4\gamma_{\nu}} +
\beta_{4\gamma_{\nu}\gamma_5} + \bar \beta_4- \bar \beta_{4\gamma_5} - \bar \beta_{4\gamma_{\nu}} +
\bar \beta_{4\gamma_{\nu}\gamma_5}.
\eea
We will use the well known form of the HMChPT lagrangian~\cite{casalbuoni}
\bea\label{L1}
&&{\cal L} = {\cal L}_{\rm light}+ {\cal L}_{\frac{1}{2}^-} + {\cal L}_{\rm ct.}\,, \nn \\
&&{\cal L}_{\rm light}=\frac{f^2}{8}\textrm{tr}\,\bigl[(\partial_\mu\Sigma^\dagger)\,
(\partial^\mu\Sigma) + \Sigma^\dagger\chi + \chi^\dagger\Sigma\,\bigr] \,,\nn\\
&& {\cal L}_{\frac{1}{2}^-} = i {\rm Tr}\left[ H_b v\negcdot D_{ba} \overline H_a\right]
+  g  {\rm Tr}\left[ H_b \gamma_\mu \gamma_5 {\bf A}^\mu_{ba} \overline H_a\right]\,,\nn\\
&&{\cal L}_{\rm ct} = k_1 {\rm Tr}\left[  \overline H_a H_b  \left( \xi  {\cal M}\xi+\xi^\dagger {\cal M}\xi^\dagger \right)_{ba}\right] + k_2{\rm Tr}\left[  \overline H_a H_a  \left( \xi  {\cal M}\xi+\xi^\dagger {\cal M}\xi^\dagger \right)_{bb}\right]\,,
\eea 
where $\overline H_a(v) = \gamma_0 H_a^\dagger (v) \gamma_0$, $g$ is the coupling of 
the pseudo-Goldstone boson to the pair of heavy-light mesons, and 
\bea
&&D^\mu_{ba}H_b 
=  \partial^\mu H_a -  H_b {1\over 2}[ \xi^\dagger \partial_\mu \xi +
\xi \partial_\mu \xi^\dagger ]_{ba}\,,\quad
{\bf A}_\mu^{ab}
= {i\over 2}[ \xi^\dagger \partial_\mu \xi -
\xi \partial_\mu \xi^\dagger ]_{ab} \;,\nn\\
&&\xi =
\sqrt{\Sigma}\,,\qquad \Sigma = \exp\left(2i\frac{\phi}{f}\right)\,,\qquad {\cal M}={\rm diag}(m_u,m_d,m_s)\,,\nn
\eea
\bea
\phi =\left(
\begin{array}{ccc}
{1\over \sqrt{2} } \pi^0  +  {1\over \sqrt{6} } \eta  &
 \pi^+  & K^+ \\
 \pi^-    &   -{1\over \sqrt{2} } \pi^0  +  {1\over \sqrt{6} } \eta  & K^0\\
K^-& \bar K^0  &-{2\over \sqrt{6} } \eta  \\
\end{array}
\right)\,,
\eea
with $f\approx 130$~MeV, and  $\chi = 2 B_0 {\cal M}$.  In the above formulae $H_a$ 
refers to either $H^{Q}_a$ or $H^{\bar Q}_a$, defined in eq.~(\ref{fieldsH}).  Note also that we distinguish between  
the trace 
over Dirac  (``Tr")  and flavor (``tr") indices.  In the chiral power counting the lagrangian ${\cal L}_{\rm light}$ in eq.~(\ref{L1})  is of ${\cal O}(p^2)$ while the rest of ${\cal L}$ is of ${\cal O}(p^1)$.  
To get the chiral logarithmic corrections to $\widetilde B_{iq}$-parameters, we should subtract twice the chiral corrections to the decay constant $\hat f_q$ from the chiral corrections to 
the four-quark operators~(\ref{bosbase}).  The former is obtained from the study of the bosonised left-handed weak current 
\bea\label{eqVA}
(V-A)_q^\mu = {i\alpha\over 2} \left\{ {\rm Tr}\left[ 
(\xi \overline H)_q \gamma^\mu(1-\gamma_5)\right]  \left( 1 + \varkappa _2 {\rm tr}{\cal M}\right) +  
 \varkappa _1 {\rm Tr}\left[  (\xi {\cal M} \overline H)_q \gamma^\mu(1-\gamma_5)\right] \right\}\,,
\eea
where $\alpha$ is the tree level decay constant in the chiral expansion, and $\varkappa _{1,2}$ are the counter-term coefficients. Together with the strong coupling $g$, these parameters are not predicted within  HMChPT. Instead, they are expected to be fixed by matching the HMChPT expressions with the results of lattice QCD for a given quantity (see reviews in ref.~\cite{sharpe}). The notation used above is the same as in ref.~\cite{QUENCH}. 
\begin{figure}
\vspace*{-0.3cm}
\begin{center}
\begin{tabular}{@{\hspace{-0.25cm}}c}
\psfrag{0-}[tc]{$\color{red}  0^-$}                                                          
\psfrag{1-}[tc]{$\color{red}  1^-$}                                                          
\psfrag{0+}[tc]{$\color{red}  0^+$}
\epsfxsize12.6cm\epsffile{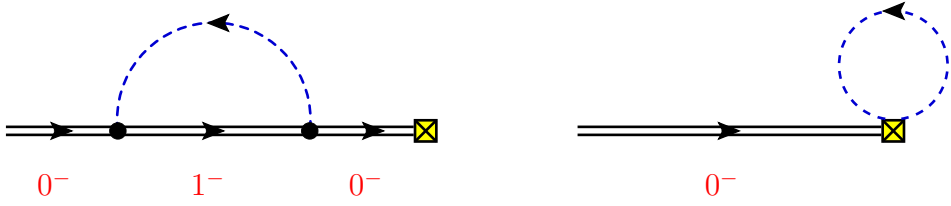}    \\
\end{tabular}
%\vspace*{-.1cm}
%%%%%%%%%%%%%%%%%%%%%%%%%%%%%%%%%%%%%%%%%%%%%%%%%%%%%%%%%%%%%%%%%%
\caption{\label{fig0}{\footnotesize\sl The diagrams which give non-vanishing chiral logarithmic corrections to the pseudoscalar heavy-light meson decay constant. The double line indicates the heavy-light meson and the dashed one the pseudo-Goldstone boson propagator. The square stands for the weak current vertex. The full dot is proportional to the coupling $g$. }}
%%%%%%%%%%%%%%%%%%%%%%%%%%%%%%%%%%%%%%%%%%%%%%%%%%%%%%%%%%%%%%%%%%
\end{center}
\end{figure}
The chiral logarithmic corrections to the decay constant come from the diagrams shown in fig.~\ref{fig0} 
\bea\label{fq}
&&\hat f_d = \alpha \left[ 1 - {1\over (4\pi f)^2 }
 \left( 
 {3\over 4}m_\pi^2\log{m_\pi^2\over \mu^2} + {1\over 2}m_K^2\log{m_K^2\over \mu^2} +{1\over 12}  m_\eta^2\log{m_\eta^2\over \mu^2}\right) \right.\nn\\
&&\hspace*{19mm} \left.+ \varkappa _1(\mu) m_d + \varkappa _2(\mu)(m_u+m_d+m_s) + {1\over 2}\delta Z_d\right]\,,\nn\\
&&\hat f_s = \alpha \left[ 1 - {1\over (4\pi f)^2 }\left( m_K^2\log{m_K^2\over \mu^2} +{1\over 3}  m_\eta^2\log{m_\eta^2\over \mu^2}\right) \right.\\
 &&\hspace*{19mm} \left.+ \varkappa _1(\mu) m_s + \varkappa _2(\mu)(m_u+m_d+m_s) + {1\over 2}\delta Z_s\right]\,,\nn
\eea
where it should be stressed that we work in the exact isospin limit ($m_u=m_d$) so that the index $d$ means either $u$- or $d$-quark. 
Only explicit in the above expressions is the term arising from the tadpole diagram (right in fig.~\ref{fig0}), whereas $Z_{d,s}$, the heavy meson field renormalization factors, come from the self energy diagram  (left in fig.~\ref{fig0}) and they read
\bea\label{Zq}
&&Z_d  = 1 - {3g^2\over (4\pi f)^2 } \left(
 {3\over 2}m_\pi^2\log{m_\pi^2\over \mu^2} + m_K^2\log{m_K^2\over \mu^2} +{1\over 6}  m_\eta^2\log{m_\eta^2\over \mu^2} \right)\nn \\
 &&\hspace{3.1cm}+ k_1(\mu) m_d + k_2(\mu)(m_u+m_d+m_s) ,\\
&&Z_s = 1 - {3 g^2\over (4\pi f)^2 }\left(2 m_K^2\log{m_K^2\over \mu^2} +{2 \over 3}  m_\eta^2\log{m_\eta^2\over \mu^2}\right) + k_1(\mu) m_s + k_2(\mu)(m_u+m_d+m_s)\ .\nn
\eea
In both eqs.~(\ref{fq}) and~(\ref{Zq}) the $\mu$ dependence in the logarithm cancels against the one in the local counter-terms. 

\begin{figure}
\vspace*{-0.3cm}
\begin{center}
\begin{tabular}{@{\hspace{-0.25cm}}c}
\psfrag{0-}[tc]{$\color{red}  0^-$}                                                          
\psfrag{1-}[tc]{$\color{red}  1^-$}                                                          
\psfrag{0+}[tc]{$\color{red}  0^+$}
\epsfxsize16.6cm\epsffile{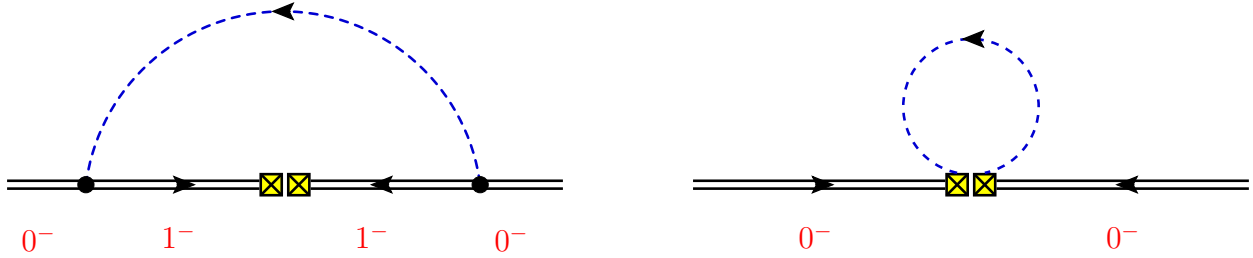}    \\
\end{tabular}
%%%%%%%%%%%%%%%%%%%%%%%%%%%%%%%%%%%%%%%%%%%%%%%%%%%%%%%%%%%%%%%%%%
\caption{\label{fig1}{\footnotesize\sl The diagrams relevant to the chiral corrections to the SM bag parameter $\widetilde B_{1q}$. In the text we refer to the left one as ``sunset", and to the right one as ``tadpole". Only the tadpole diagram gives a non-vanishing contribution to the bag parameters $\widetilde B_{2,4q}$.} }
%%%%%%%%%%%%%%%%%%%%%%%%%%%%%%%%%%%%%%%%%%%%%%%%%%%%%%%%%%%%%%%%%%
\end{center}
\end{figure}
With these ingredients in hands it is now easy to deduce that the only diagrams which contribute to the SM bag parameter,  
$\widetilde B_{1q}$, are the two shown in fig.~\ref{fig1}.  They arise from the two terms  in 
$\widetilde O_1 = 4\widehat \beta_1 [ (\xi \bar P^\ast_\mu)_q (\xi \bar P^{\ast \mu})_q  + (\xi \bar P)_q (\xi \bar P)_q]$ and  yield 
\bea
{\rm ``sunset"} :&& 4\widehat \beta_1 {3g^2\over (4\pi f)^2 }\sum_i (t^i_{qq})^2 m_{i}^2\log{m_{i}^2\over \mu^2}\,,\nn\\
{\rm ``tadpole"} : &-&4\widehat \beta_1  {1\over (4\pi f)^2} \sum_i (t^i_{qq})^2 m_{i}^2\log{m_{i}^2\over \mu^2}\,,
\eea
respectively,  where $t^i$ are the SU(3) generators and $m_i$ masses of the pseudo-Goldstone bosons. The SM bag parameters now read
\bea \label{B1}
&&\widetilde B_{1d}=\widetilde B_1^{\rm Tree}\left[ 1 - {1-3g^2\over (4\pi f)^2}\left(
{1\over 2}m_\pi^2\log{m_\pi^2\over \mu^2} + {1\over 6}m_\eta^2\log{m_\eta^2\over \mu^2} \right)\right.\nn\\
 &&\hspace{4.2cm} \biggl.+ b_1(\mu) m_d +b_1^\prime(\mu) (m_u+m_d+m_s) \biggr]\,,\\
&&\widetilde B_{1s}=\widetilde B_1^{\rm Tree}\left[ 1 - {1-3g^2\over (4\pi f)^2} 
 {2\over 3}m_\eta^2\log{m_\eta^2\over \mu^2} + b_1(\mu) m_s +b_1^\prime(\mu) (m_u+m_d+m_s)
\right]\,,\nn
\eea
where we also wrote the counter-term contributions and, for short, we wrote $\widetilde B_1^{\rm Tree}=3\widehat \beta_1/2\alpha^2$.  The above results agree  with the ones 
presented in refs.~\cite{grinstein,eeg}, in which the pion loop contribution was left out, and with the ones  recently presented in ref.~\cite{dlin}.

As for the bag parameters $\widetilde B_{2q}$ and $\widetilde B_{4q}$ we obtain
\bea \label{B24}
\widetilde B_{2,4 d}&=&\widetilde B_{2,4}^{\rm Tree}\left[ 1 + {3 g^2 Y\mp 1\over (4\pi f)^2}\left(
{1\over 2}m_\pi^2\log{m_\pi^2\over \mu^2} + {1\over 6}m_\eta^2\log{m_\eta^2\over \mu^2} \right) \right. \nn\\ 
&& \Biggl. \hspace*{11mm} + b_{2,4}(\mu) m_d +b_{2,4}^\prime(\mu) (m_u+m_d+m_s) \Biggr] \,,\\
\widetilde B_{2,4s}&=&\widetilde B_{2,4}^{\rm Tree}\left[ 1 +{2\over 3} {3 g^2 Y \mp 1\over (4\pi f)^2} 
 m_\eta^2\log{m_\eta^2\over \mu^2} + b_{2,4}(\mu) m_s +b_{2,4}^\prime(\mu) (m_u+m_d+m_s)\right]
,\nn
\eea
where  $\widetilde B_2^{\rm Tree}=12\widehat \beta_2/4\alpha^2$, $\widetilde B_4^{\rm Tree}=12\widehat \beta_2/7\alpha^2$
$Y=(\widehat \beta^*_{2,4}/\widehat \beta_{2,4})$, with 
$\widehat \beta^\ast_2 = \beta_{2\gamma_{\nu}} +
\beta_{2\gamma_{\nu}\gamma_5} + 4 \beta_{2\sigma_{\nu\rho}}$, and $ \widehat
\beta^\ast_4 =  \beta_{4\gamma_{\nu}} - \beta_{4\gamma_{\nu}\gamma_5}+ \bar \beta_{4\gamma_{\nu}} - \bar \beta_{4\gamma_{\nu}\gamma_5}$. We checked that our results agree with those presented in ref.~\cite{dlin} where also the partially quenched theory has been considered. In our paper we refer 
only to the full (unquenched) theory. 

\section{Impact of the $1/2^+$-mesons}

In this section we examine the impact of the heavy-light mesons belonging to the 
$1/2^+$ doublet when propagating in the loops onto the chiral logarithmic corrections derived in the previous section.  We first extend the lagrangian by adding to eq.~(\ref{L1}) the following terms~\cite{casalbuoni}:
\bea\label{Ls}
{\cal L}_{\frac{1}{2}^+}& =& - {\rm Tr}\left[ S_b(i v\negcdot D_{ba} + \Delta_S) \overline S_a\right]
+  \tilde g  {\rm Tr}\left[ S_b \gamma_\mu \gamma_5
{\bf A}^\mu_{ba} \bar S_a\right]\,,\nn\\
{\cal L}_{\rm mix} & =&  h {\rm Tr}\left[ S_b \gamma_\mu \gamma_5 {\bf A}^\mu_{ba} \overline H_a\right] +
{\rm h.c.}\,,\\
{\cal L}_{\rm ct} & =&   \widetilde k_1 {\rm Tr}\left[\overline S_a S_b 
\left( \xi  {\cal M}\xi+\xi^\dagger {\cal M}\xi^\dagger \right)_{ba}\right] + \widetilde k_2 {\rm Tr}\left[ \overline S_a S_a \left( \xi  {\cal M}\xi+\xi^\dagger {\cal M}\xi^\dagger \right)_{bb}\right]\nn\\
& +&  k_1^\prime {\rm Tr}\left[\overline H_a S_b 
\left( \xi  {\cal M}\xi+\xi^\dagger {\cal M}\xi^\dagger \right)_{ba}\right] + k_2^\prime {\rm Tr}\left[ \overline H_a S_a \left( \xi  {\cal M}\xi+\xi^\dagger {\cal M}\xi^\dagger \right)_{bb}\right]+
{\rm h.c.}\,,\nn
\eea 
where the fields of  the scalar ($P_0$) and the axial
($P^{\ast}_{1\ \mu}$) mesons are organised in a superfield
\bea
S_q(v) = {1 +  \vdir \over 2} \left[ P^{\ast}_{1\ \mu} (v)\gamma_\mu \gamma_5 - P_0 (v)
\right]_q\;,\qquad
\overline S_q(v) = \gamma_0 S_q^\dagger (v) \gamma_0\,.
\eea
$\tilde g$ is the coupling of the $P$-wave Goldstone boson to the pair of $1/2^+$  heavy-light mesons, and $h$ is the coupling of the $S$-wave Goldstone boson to the heavy-light mesons, one of which belongs to $1/2^-$ and the other to $1/2^+$ doublet.  Before including the $1/2^+$ doublet we were free to set $\Delta =0$ because all the chiral loop divergences are cancelled by ${\cal O}(m_q)$ counter-terms in the static heavy quark limit. Once the $1/2^+$ doublet is included, the mass difference between the  $1/2^+$ and  $1/2^-$  states ($\Delta_S\approx 400$~MeV) must be included in the lagrangian, but since it does not vanish in the chiral nor in the heavy quark limit it is of ${\cal O}(p^0)$ in the chiral power counting~(see also ref.~\cite{springer} where, in addition to the static heavy quark limit, the  chiral power counting is discussed also when the $1/m_Q$-corrections are included).

\subsection{Decay constants}

\begin{figure}
\vspace*{-0.3cm}
\begin{center}
\begin{tabular}{@{\hspace{-0.25cm}}c}
\psfrag{0-}[tc]{$\color{red}  0^-$}                                                          
\psfrag{1-}[tc]{$\color{red}  1^-$}                                                          
\psfrag{0+}[tc]{$\color{red}  0^+$}
\epsfxsize12.6cm\epsffile{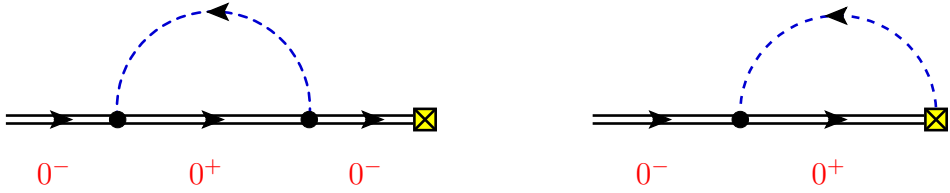}    \\
\end{tabular}
%\vspace*{-.1cm}
%%%%%%%%%%%%%%%%%%%%%%%%%%%%%%%%%%%%%%%%%%%%%%%%%%%%%%%%%%%%%%%%%%
\caption{\label{fig3}{\footnotesize\sl In addition  to the diagrams shown 
in fig.~\ref{fig0}, these two diagrams  contribute the loop corrections to the 
pseudoscalar meson decay constant after the $1/2^+$ mesons are included in HMChPT. The full dot in these graphs is proportional to the coupling $h$. }}
%%%%%%%%%%%%%%%%%%%%%%%%%%%%%%%%%%%%%%%%%%%%%%%%%%%%%%%%%%%%%%%%%%
\end{center}
\end{figure}

Beside the lagrangian, the $1/2^+$ mesons contribution should also be added to the left vector current~(\ref{eqVA}), which now reads 
\bea\label{VAscalar}
(V-A)_q^\mu &=& {i\alpha\over 2} {\rm Tr}\left[ 
(\xi \overline H)_q \gamma^\mu(1-\gamma_5)\right]  + 
 {i \alpha^+ \over 2} {\rm Tr}\left[ 
(\xi \overline S)_q \gamma^\mu(1-\gamma_5)\right]   \nn\\ &+&
{i\alpha \over 2}  \varkappa _1 {\rm Tr}\left[  (\xi {\cal M} \overline H)_q \gamma^\mu(1-\gamma_5)\right]   +{i \alpha^+  \over 2} \widetilde \varkappa _1 {\rm Tr}\left[  (\xi {\cal M} \overline S)_q \gamma^\mu(1-\gamma_5)\right]   \nn\\ &+&
{i\alpha \over 2}  \varkappa _2 {\rm Tr}\left[  (\xi  \overline H)_q \gamma^\mu(1-\gamma_5)\right] {\rm tr}{\cal M}  +{i \alpha^+  \over 2} \widetilde \varkappa _2 {\rm Tr}\left[  (\xi \overline S)_q \gamma^\mu(1-\gamma_5)\right]{\rm tr}{\cal M} \,,
\eea
where $\alpha^+ $ is the coupling of one of the $1/2^+$  
mesons to the weak left current, and $\widetilde \varkappa _{1,2}$ are the coefficients of two new counter-terms.  
From the recent lattice results reported in  ref.~\cite{craig}  we extract,  $\alpha^+ /\alpha=1.1(2)$. In other words, at least in the static heavy quark mass limit,  the weak current coupling of $1/2^+$ mesons is not suppressed with respect to the $1/2^-$ ones. Since we focus on the pseudoscalar meson decay constant, it should be clear that only the scalar meson from the $1/2^+$ doublet can propagate in the loop.  
The diagrams that give non-vanishing contributions are shown in fig.~\ref{fig3} and the corresponding expressions now read
\bea
&&Z_{q}=1 +  {t^i_{qa}t^{i\dagger}_{aq} \over  (4\pi f)^2}  
\left\{ 3 g^2 \lim_{x\to 0} {d\over dx}[xJ_1(m_i^2,x)] - h^2 \Bigl[
J_1(m_i^2,\Delta_S) + J_2(m_i^2,\Delta_S) \Bigr.\right. \nn\\
&& \hspace*{7.16cm}\left.\Bigl. 
 +  \Delta_S {d\over d\Delta_S}\bigl(
 J_1(m_i^2,\Delta_S) + J_2(m_i^2,\Delta_S) \bigr) \Bigr] \right\} \,,\nn \\
&&\hat f_{q} = \alpha \left\{ 
1 + { t^i_{qa}t^{i\dagger}_{aq} \over 2 (4\pi f)^2}  \biggl[ 
3 g^2 \lim_{x\to 0} {d\over dx}[xJ_1(m_i^2,x)] - I_1(m_i^2) 
-h^2 \biggl(
J_1(m_i^2,\Delta_S) + J_2(m_i^2,\Delta_S) \biggr.\biggr.\right.
\nn\\
&&\hspace*{.6cm} \left.\left.\biggl.+ \Delta_S {d\over d\Delta_S}\left(
 J_1(m_i^2,\Delta_S) + J_2(m_i^2,\Delta_S) \right)
\biggr)
- 2 h{\alpha^+ \over \alpha}\left(
I_1(m_i^2) + I_2(m_i^2,\Delta_S) \right) 
\right]\right\}\,,
\eea
where the summation over ``$i$" is implicit, and we omit the counter-term contributions to make the expressions simpler. We stress that in all our formulae, the terms of ${\cal O}(p^3)$ and higher in the chiral power counting are neglected. The integrals $I_{1,2}$ and $J_{1,2}$ are the same as the ones  
used in ref.~\cite{QUENCH} and can be found in the appendix of that paper. The terms proportional to $h^2$ comes from the inclusion of the left diagram shown in fig.~\ref{fig3}, while the last term in the decay constant (the one proportional to $h$) comes from the right graph in fig.~\ref{fig3}. Obviously 
such terms were absent before including the scalar mesons. Notice also that 
\bea\label{limits0}
 \lim_{x\to 0} {d\over dx}[xJ_1(m_\pi^2,x)] = - m_\pi^2\log{m_\pi^2\over \mu^2} \,.
\eea
 When $\Delta_S > m_i$, which in our case is 
true for the pion mass, we can expand around $m_\pi^2=0$ and obtain
\bea\label{limits1}
  J_1(m_\pi^2,\Delta_S) + J_2(m_\pi^2,\Delta_S) =&& \nn \\ 
  I_1(m_\pi^2) + I_2(m_\pi^2,\Delta_S) 
&\to& 2 \Delta_S^2 ( 1- \log{4\Delta_S^2\over \mu^2}) + m_\pi^2 ( 1+ \log{4\Delta_S^2\over \mu^2}) + \dots\,,\nn\\
 -  \Delta_S {d\over d\Delta_S}\bigl[ J_1(m_\pi^2,\Delta_S) + J_2(m_\pi^2,\Delta_S) \bigr] &\to& 4\Delta_S^2 \log{4\Delta_S^2\over \mu^2}- 2 m_\pi^2 + \dots\,, 
\eea
where the dots stand for higher powers in $m_\pi^2$. In other words, the presence of the nearby $1/2^+$ state does not affect the pionic logarithmic behavior of the decay constant. It does, however, affect the kaon and $\eta$-meson loops because those states are heavier than $\Delta_S$ ($m_\pi < \Delta_S \lesssim m_K < m_\eta$) and the coefficients of their logarithms, although still predictions of this approach,  cease to be numerically relevant because those logarithms are competitive in size with the terms proportional to $\Delta_S^2 \log(4\Delta_S^2/ \mu^2)$, as indicated in eq.~(\ref{limits1}). 
Stated equivalently, the relevant chiral logarithmic corrections are those coming from the $SU(2)_L\otimes SU(2)_R \to SU(2)_V$ theory, and the pseudoscalar decay constant reads
\bea\label{fB-correct}\boxed{{\phantom{\Large{l}}}\raisebox{.4cm}{\phantom{\Large{j}}}
\quad \hat f_{q} = \alpha \left[ 1-  {1+3g^2\over 2 (4\pi f)^2} {3\over 2} m_\pi^2 \log{m_\pi^2\over \mu^2} + c_f(\mu) m_\pi^2 \right]\quad
}\,,\eea 
where $c_f(\mu)$ stands for the combination of the counter-term coefficients considered in the 
previous section.~\footnote{More specifically, $2B_0 c_f(\mu) + \displaystyle{{3h^2 \over 4(4\pi f)^2}} \left[ 
3+\log(4\Delta_S^2/\mu^2)\right]+\displaystyle{{3 h \alpha^+  \over 2\alpha (4\pi f)^2}} [1+\log(4\Delta_S^2/\mu^2)] = \frac{1}{2}k_1(\mu)+ \frac{1}{2}k_1^\prime (\mu)+k_2(\mu)+k_2^\prime (\mu)+\varkappa _1(\mu)+2\varkappa _2(\mu)$, where we use the Gell-Mann--Oakes--Renner formula, $m_\pi^2 = 2 B_0 m_d$. The exact isospin symmetry ($m_u=m_d$) is assumed throughout this work. }  
At this point we also note that we checked that the chiral logarithms in the scalar heavy-light meson decay constant, which has recently been computed on the lattice in ref.~\cite{craig}, are the same as for the pseudoscalar meson, 
with the coupling $g$ being replaced by $\widetilde g$, i.e.,
\bea\label{su2f}\boxed{{\phantom{\Large{l}}}\raisebox{.4cm}{\phantom{\Large{j}}}
\quad\hat  f_{q}^+ = \alpha^+ \left[ 1-  {1+3\widetilde g^2\over 2 (4\pi f)^2} {3\over 2} m_\pi^2 
\log{ m_\pi^2\over \mu^2} +  c^+_f(\mu) m_\pi^2 \right]\quad 
}\,.\eea
Since $\widetilde g^2/g^2 \approx 1/9$~\cite{bloss}, the deviation from the linear behavior in $m_\pi^2$ 
is less pronounced for  $\hat f_{q}^+$ than it is for $\hat f_{q}$.  
Finally, it should be emphasized that the 
counter-term coefficients relevant to the $SU(2)_V$-theory, namely $c_f^{(+)}(\mu)$   
in eqs.~(\ref{fB-correct},\ref{su2f}), are not the same as those in $SU(3)_V$.

\subsection{Bag parameters}

In this subsection we show that the situation with the bag parameters is similar to the one with decay constant, namely the pion loop chiral logarithmis remain unchanged when the nearby scalar meson is included in HMChPT. 
To that end, besides eq.~(\ref{Ls}), we should include the contributions of $1/2^+$-mesons to the  operators~(\ref{bosbase}). Generically the operators $\widetilde{\cal O}_{1,2,4}$ now become
\bea\label{eq-9}
\widetilde O_{1}&= & \sum_X \beta_{1 X} {\rm Tr}\left[ \left(\xi \overline  H^Q\right)_q \gamma_{\mu }(1-\gamma_5) X\right]
 {\rm Tr}\left[ \left(\xi H^{\bar Q}\right)_q \gamma^{\mu }(1-\gamma_5) X \right] \nonumber \\
&&\hspace*{5mm}  +  
 \beta_{1 X}^\prime  \left\{ 
 {\rm Tr}\left[ \left(\xi \overline  H^Q\right)_q \gamma_{\mu }(1-\gamma_5) X  \right]
 {\rm Tr}\left[ \left(\xi S^{\bar Q}\right)_q  \gamma^{\mu }(1-\gamma_5) X  \right]   +{\rm h.c.}\right\} \nonumber \\
&&\hspace*{5mm}+  \beta_{1 X}^{\prime\prime}  {\rm Tr}\left[ \left(\xi \overline  S^Q\right)_q \gamma_{\mu }(1-\gamma_5) X \right]
 {\rm Tr}\left[ \left(\xi S^{\bar Q}\right)_q  \gamma^{\mu }(1-\gamma_5) X\right] \,,
\eea
where $\beta_{1X}^{\prime}$ are the couplings of the operator  $\widetilde O_{1}$ 
to both $1/2^-$ and $1/2^+$ mesons, while $\beta_{1X}^{\prime\prime}$ come from the coupling to the 
$1/2^+$ mesons only.  Similarly, the operators $\widetilde O_{2,4}$ now read:
\bea\label{eq-92}
\widetilde O_{2}&= & \sum_X \beta_{2 X} {\rm Tr}\left[ \left(\xi \overline  H^Q\right)_q  (1-\gamma_5) X\right]
 {\rm Tr}\left[ \left(\xi H^{\bar Q}\right)_q  (1-\gamma_5) X \right] \nonumber \\
&&\hspace*{5mm}  +  
 \beta_{2 X}^\prime  \left\{ 
 {\rm Tr}\left[ \left(\xi \overline  H^Q\right)_q  (1-\gamma_5) X  \right]
 {\rm Tr}\left[ \left(\xi S^{\bar Q}\right)_q   (1-\gamma_5) X  \right]   +{\rm h.c.}\right\} \nonumber \\
&&\hspace*{5mm}+  \beta_{2 X}^{\prime\prime}  {\rm Tr}\left[ \left(\xi \overline  S^Q\right)_q  (1-\gamma_5) X \right]
 {\rm Tr}\left[ \left(\xi S^{\bar Q}\right)_q  (1-\gamma_5) X\right] \,, 
\eea
\bea\label{eq-93}
\widetilde O_{4}
&= & \sum_X \beta_{4 X} {\rm Tr}\left[ \left(\xi \overline  H^Q\right)_q  (1-\gamma_5) X\right]
 {\rm Tr}\left[ \left(\xi^\dagger H^{\bar Q}\right)_q  (1+\gamma_5) X \right] \nonumber \\
&&\hspace*{5mm}  + \bar \beta_{4 X} {\rm Tr}\left[ \left(\xi   H^{\bar Q}\right)_q  (1-\gamma_5) X\right]
 {\rm Tr}\left[ \left(\xi^\dagger \overline H^Q\right)_q  (1+\gamma_5) X \right] \nonumber \\
&&\hspace*{5mm}  +  
 \beta_{4 X}^\prime  \left\{ 
 {\rm Tr}\left[ \left(\xi \overline  H^Q\right)_q  (1-\gamma_5) X  \right]
 {\rm Tr}\left[ \left(\xi^\dagger S^{\bar Q}\right)_q   (1+\gamma_5) X  \right]   +{\rm h.c.}\right\} \nonumber \\
&&\hspace*{5mm}  +  
\bar \beta_{4 X}^\prime  \left\{ 
 {\rm Tr}\left[ \left(\xi  H^{\bar Q}\right)_q  (1-\gamma_5) X  \right]
 {\rm Tr}\left[ \left(\xi^\dagger \overline S^Q\right)_q   (1+\gamma_5) X  \right]   +{\rm h.c.}\right\} \nonumber \\
&&\hspace*{5mm}+  \beta_{4 X}^{\prime \prime} {\rm Tr}\left[ \left(\xi \overline  S^Q\right)_q  (1-\gamma_5) X\right]
 {\rm Tr}\left[ \left(\xi^\dagger S^{\bar Q}\right)_q  (1+\gamma_5) X \right] \nonumber \\
&&\hspace*{5mm}  + \bar \beta_{4 X}^{\prime \prime}  {\rm Tr}\left[ \left(\xi   S^{\bar Q}\right)_q  (1-\gamma_5) X\right]
 {\rm Tr}\left[ \left(\xi^\dagger \overline S^Q\right)_q  (1+\gamma_5) X \right] \,.
\eea
After evaluating the traces in eqs.~(\ref{eq-9}), and keeping in mind that the external states are the pseudoscalar mesons, we have
\bea
\widetilde O_1&= & 4\widehat \beta_1\left[   \left(\xi \overline  P^{Q \ast \mu} \right)_q  \left(\xi P^{\bar Q\ \ast}_\mu \right)_q
 +   \left(\xi \overline  P^Q \right)_q  \left(\xi P^{\bar Q}  \right)_q \right]  \nn\\
&& + 4\widehat \beta_1^\prime \left[ 
    \left(\xi \overline  P^Q \right)_q  \left(\xi P^{\bar Q}_0  \right)_q  +\left(\xi \overline  P^Q_0 \right)_q  \left(\xi P^{\bar Q}  \right)_q \right] + 4\widehat \beta_1^{\prime\prime}  \left(\xi \overline  P^Q_0 \right)_q  \left(\xi P^{\bar Q}_0  \right)_q  \,,  
\eea
where $\widehat \beta_1^{(\prime,\prime\prime)}$ have forms analogous to the ones written 
in eq.~(\ref{betas}).  In addition, in eqs.~(\ref{eq-9},\ref{eq-92},\ref{eq-93}), the fields  $ S^{Q}_q$ and $ S^{\bar Q}_q$ are defined in a way similar to eq.~(\ref{fieldsH}). 
\begin{figure}
\vspace*{-0.3cm}
\begin{center}
\begin{tabular}{@{\hspace{-0.25cm}}c}
\psfrag{0-}[cc]{$\color{red}  0^-$}                                                          
\psfrag{1-}[cc]{$\color{red}  1^-$}                                                          
\psfrag{0+}[cc]{$\color{red}  0^+$}                                                          
\psfrag{1-0+}[cc]{\scriptsize{$ \color{red}{\quad 1^-, 0^+}$}}
\epsfxsize15.9cm\epsffile{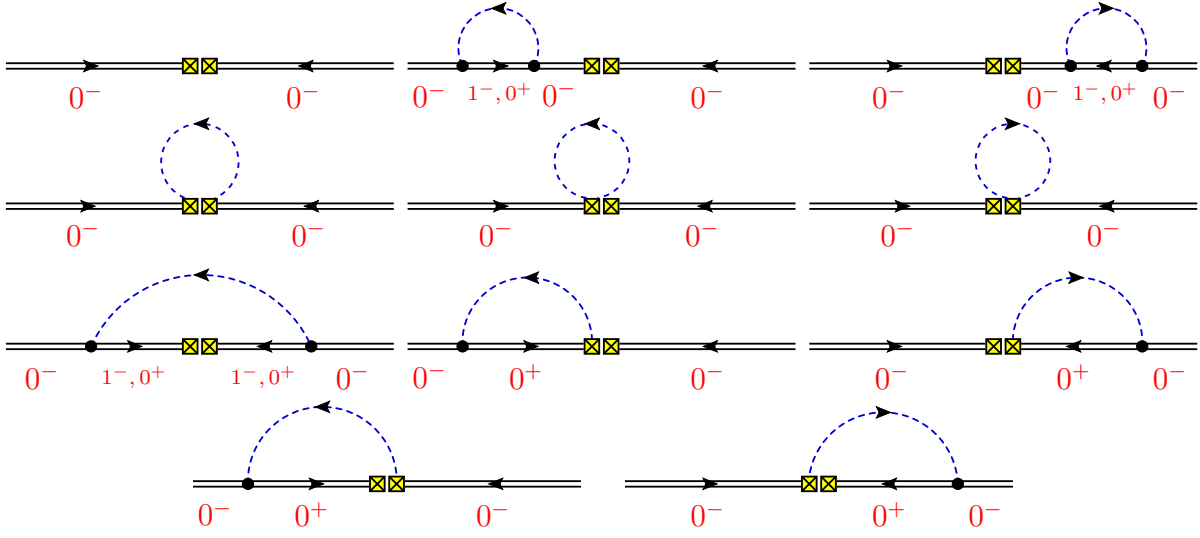}    \\
\end{tabular}
%\vspace*{-.1cm}
%%%%%%%%%%%%%%%%%%%%%%%%%%%%%%%%%%%%%%%%%%%%%%%%%%%%%%%%%%%%%%%%%%
\caption{\label{fig4}{\footnotesize\sl All diagrams which enter in the calculation of the chiral corrections to the operators 
$\langle \widetilde {\cal O}_{1,2,4}\rangle$.  Note that we indicate when either the vector or scalar meson can propagate in the loop by $1^-,0^+$, in contrast to the cases when only the scalar ($0^+$) meson can couple. }}
%%%%%%%%%%%%%%%%%%%%%%%%%%%%%%%%%%%%%%%%%%%%%%%%%%%%%%%%%%%%%%%%%%
\end{center}
\end{figure}
The corresponding tree and the 1-loop chiral diagrams are shown in fig.~\ref{fig4}.  Since the couplings of the four-quark operators to the scalar meson are proportional to $\beta_i^{\prime,\prime\prime}$ and of the pseudoscalar decay constant to $\alpha^+ $, the cancellation between 
the chiral loop corrections in the operators  $\widetilde O_{i}$ and in the decay constant is not automatic. For that reason, instead of writing the chiral logarithmic corrections to the bag-parameter, we will write them for the  full operator, namely
\bea\label{eqO1}
\widetilde B_{1q} \hat f_q^2 &=& {3\over 2}\widehat \beta_1
 \left\{ 
1 - { t^i_{qa}t^{i\dagger}_{aq} \over 2 (4\pi f)^2}  \biggl[ 
-6 g^2 \lim_{x\to 0} {d\over dx}[xJ_1(m_i^2,x)] + 2 I_1(m_i^2)\biggr.\right.
\nn\\
&&+2h^2 \bigl(
J_1(m_i^2,\Delta_S) + J_2(m_i^2,\Delta_S)+ \Delta_S {d\over d\Delta_S}\left(
 J_1(m_i^2,\Delta_S) + J_2(m_i^2,\Delta_S) \right)
\biggr)
\nn\\
&& \biggl. +4 h{\widehat \beta_1^\prime\over \widehat  \beta_1}\left(
I_1(m_i^2) + I_2(m_i^2,\Delta_S) \right) 
 \biggr]  - { t^i_{qq}t^{i\dagger}_{qq} \over 2 (4\pi f)^2} \biggl[ 
6 g^2 \lim_{x\to 0} {d\over dx}[xJ_1(m_i^2,x)]  +   2 I_1(m_i^2)  \biggr.
 \nn\\
&& \biggl.  \biggl.
+ 4 h {\widehat \beta_1^{\prime}\over \widehat \beta_1}\left(
I_1(m_i^2) + I_2(m_i^2,\Delta_S) \right) + 
h^2 {\widehat \beta_1^{\prime\prime}\over \widehat \beta_1} \sum_{k=1,2;s=\pm 1}J_k(m_i^2,s\Delta_S) 
\biggr]\biggr\}\,,
\eea
To keep the above expression simpler we do not write  
the counter-terms  since their structure remains the same as before, i.e., a constant times $m_q$ and an another  constant times ${\rm tr}{\cal M}$. 
The similar formulae for $\widetilde B_{2,4q}\hat f_q^2$ are lengthy and we will not write 
them explicitely. For the point that we want to make in this section it is enough to 
consider eq.~(\ref{eqO1}) because in the expressions for $\widetilde B_{2,4q}\hat f_q^2$ the loop functions $I_{1,2}$ and
$J_{1,2}$ occur in the same form as in eq.~(\ref{eqO1}). 
In  the evaluation of the sunset diagrams we used the standard simplification-trick
\bea
{1\over (vp - \Delta)(vp - \Delta^\prime)} = {1\over \Delta -\Delta^\prime}\left( {1\over vp - \Delta} - {1\over vp - \Delta^\prime}\right)\,.
\eea

We now turn to the case $m_\pi \ll \Delta_S$ and study the behavior of eq.~(\ref{eqO1}) around 
$m_\pi^2\to 0$.  In addition to the limits discussed in eq.~(\ref{limits1}), when dealing with 
the integrals in the last line of eq.~(\ref{eqO1})  
we shall proceed similarly to what has been done in  ref.~\cite{jernej}, namely  we expand the integrand in 
$E_\pi/\Delta_S$  and write
\bea\label{limits2}
\sum_{k=1,2;s=\pm 1}J_k(m_\pi^2,s\Delta_S) &=& - 2 (4\pi)^2 v_\mu v_\nu \times
 i \mu^\epsilon  \int {d^{4-\epsilon} p
\over (2\pi)^{4-\epsilon} }{p^\mu p^\nu\over (p^2-m_\pi^2) [\Delta_S^2 - (vp)^2] }\nn\\
&=& - {2 (4\pi^2)\over \Delta_S^2} v_\mu v_\nu \left[ 
 i \mu^\epsilon  \int {d^{4-\epsilon} p
\over (2\pi)^{4-\epsilon} }{p^\mu p^\nu\over p^2-m_\pi^2  } + {\cal O}(1/\Delta_S^2)
\right]\nn\\
&=& -{m_\pi^2\over 2\Delta_S^2} (4\pi^2) I_1(m_\pi^2) + \dots \to -{m_\pi^4\over 2\Delta_S^2}\log{m_\pi^2\over \mu^2} + \dots,
\eea
where the ellipses stand for the terms of higher order in $m_\pi^2/\Delta_S^2$. Note, however, that even the leading  term is already of higher order in the chiral expansion and thus the terms proportional to $h$ in eq.~(\ref{eqO1}) do not affect the leading chiral logarithmic corrections. 

On the basis of the above discussion and eqs.~(\ref{limits0},\ref{limits1}) we see that after expanding  eq.~(\ref{eqO1})   around $m_\pi^2=0$, the leading chiral logarithms arising from the pion loops remain  unchanged even when the coupling to the scalar meson is included 
in the loops. On the other hand,  as discussed in the previous subsection, 
the logarithms arising from the kaon and the $\eta$-meson are competitive in size with those arising from the coupling to the heavy-light scalar meson, which is the consequence of the  smallness of $\Delta_S$. 
Therefore, like for the decay constants,  the relevant chiral expansion 
is the one derived in the $SU(2)_L \otimes SU(2)_R \to SU(2)_V$ theory, i.e.,
\bea 
\widetilde B_{1q} \hat f_q^2 &=& \widetilde B_{1}^{\rm Tree} \alpha^2
 \left[ 
1 - {3 g^2 +2 \over  (4\pi f)^2} m_\pi^2\log{m_\pi^2\over \mu^2} + c_{{\cal O}_1}(\mu)m_\pi^2\right]\,,\nn\\ 
\widetilde B_{2,4q} \hat f_q^2 &=&  \widetilde B_{2,4}^{\rm Tree} \alpha^2
 \left[ 
1 - {3g^2(3 -Y) +3\pm 1 \over 2 (4\pi f)^2} m_\pi^2\log{m_\pi^2\over \mu^2} 
+ c_{{\cal O}_{2,4}}(\mu)m_\pi^2\right] \,,
\eea
or by using eq.~(\ref{fB-correct}), for the bag parameters we obtain
\bea\label{B1-correct}\boxed{{\phantom{\Large{l}}}\raisebox{.4cm}{\phantom{\Large{j}}}
\quad \widetilde B_{1q}  =\widetilde B_1^{\rm Tree} 
 \left[ 
1 - {1-3 g^2  \over 2 (4\pi f)^2} m_\pi^2\log{m_\pi^2\over \mu^2} + c_{{B}_1}(\mu)m_\pi^2\right]\quad   
}\,,\eea
\bea\label{B24-correct}
 \boxed{{\phantom{\Large{l}}}\raisebox{.4cm}{\phantom{\Large{j}}}
\quad \widetilde B_{2,4q}   =  \widetilde B_{2,4}^{\rm Tree}  \left[ 
 1 + {3 g^2 Y \mp 1 \over 2 (4\pi f)^2} m_\pi^2\log{m_\pi^2\over \mu^2} 
 + c_{{B}_{2,4}}(\mu)m_\pi^2\right] \quad}\,,
\eea
which coincide with the pion loop contributions shown in eqs.~(\ref{B1}) and (\ref{B24}), as they should.

\section{Relevance to the analyses of the lattice QCD data} 

It should be stressed that the consequence of the discussion in the previous section 
is mainly important to the phenomenological approaches in which the sizable kaon and $\eta$-meson logarithmic corrections  are taken as predictions, whereas the counter-term coefficients are fixed by 
matching to large $N_c$ expansion or some other model. 
We showed that the contributions of the nearby heavy-light scalar states are competitive in size and thus they cannot 
be ignored nor separated from the discussion of the kaon and/or $\eta$-meson loops. 

\begin{figure}
\vspace*{-0.3cm}
\begin{center}
\begin{tabular}{@{\hspace{-0.25cm}}c}
\epsfxsize8.6cm\epsffile{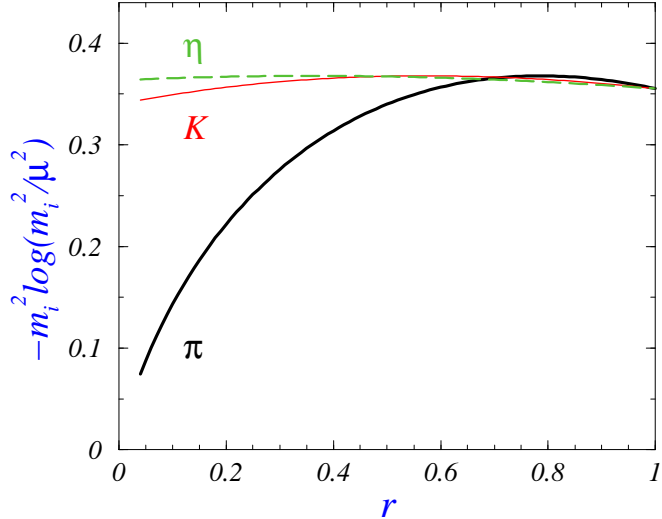}    \\
\end{tabular}
\vspace*{-.1cm}
%%%%%%%%%%%%%%%%%%%%%%%%%%%%%%%%%%%%%%%%%%%%%%%%%%%%%%%%%%%%%%%%%%
\caption{\label{fig5}{\footnotesize\sl Typical chiral logarithmic contributions $-m_i^2\log(m_i^2/\mu^2)$ are shown for pion, kaon and $\eta$ as a function of $r=m_{d}/m_s$, with $m_s$ fixed to its physical value, and $\mu=1$~GeV.} }
%%%%%%%%%%%%%%%%%%%%%%%%%%%%%%%%%%%%%%%%%%%%%%%%%%%%%%%%%%%%%%%%%%
\end{center}
\end{figure}
In the extrapolation of the lattice data, instead, this is not a problem because the kaon and the $\eta$-meson loops essentially do not alter the quark mass dependence, whereas the important nonlinearity comes from the pion chiral loops. As an illustration, in fig.~\ref{fig5} we plot the  typical chiral logarithm, 
$-m_i^2\log(m_i^2/\mu^2)$, as a function of $r=m_{d}/m_s$ which appear in the Gell-Mann--Oakes--Renner formulae, 
\bea \label{gell-mann}
m_\pi^2 = 2B_0 m_s  r \;,
\quad 
m_K^2 = 2 B_0 m_s {r + 1\over 2}\;,
\quad m_\eta^2 = 2B_0m_s {r+2\over 3}\;,
\eea
with  $2 B_0 m_s = 2 m_K^2-m_\pi^2=0.468\ \gev^2$.  Thus the fact that the nearby scalar heavy-light mesons do not spoil the pion logarithmic corrections to the decay constants and the bag-parameters is most welcome from the lattice practitioners' point of view, because the formulae derived in HMChPT can still (and should) be used to guide the chiral extrapolations of the lattice results, albeit for the pion masses lighter than $\Delta_S$.

\section{Conclusion}

In this paper we revisited the computation of the $\bbarq$ mixing amplitudes in the framework 
of HMChPT.  Besides the SM bag parameter, we also provided the expressions for the chiral 
logarithmic correction to the SUSY bag parameters. 
More importantly, we study the impact of the near scalar mesons to the predictions derived in HMChPT in which these 
contributions were previously ignored. We showed that while the corrections due to the nearness of the scalar mesons 
are competitive in size with the kaon and $\eta$ meson loop corrections, they do not alter the pion chiral logarithms.  
In other words the valid (pertinent) ChPT expressions for the quantities discussed in this paper are those involving pions only.
This is of major importance for the chiral extrapolations of the results obtained from the QCD simulations on 
the lattice, because precisely the pion chiral logarithms provide the important guidance in those extrapolations as long as $m_\pi \ll \Delta_S$. 
The corresponding useful formulas are given in eqs.~(\ref{fB-correct},\ref{B1-correct},\ref{B24-correct}).
As a  side-result we verified that the chiral logarithmic corrections to the scalar meson decay constant 
are the same as to the pseudoscalar one, modulo replacement $g\to \widetilde g$ (c.f. eq.~(\ref{su2f})).

\section*{Acknowledgement} 
It is a pleasure to thank S.~Descotes-Genon  and J.O.~Eeg for discussions. The financial support through the PAI project 
{\sl ``Proteus"} is kindly acknowledged. The work of S.F. and  J.K. was supported in part by 
the Slovenian Research Agency.

%\newpage 


\begin{thebibliography}{99}
{ \footnotesize

\bibitem{Albajar:1986it}
  C.~Albajar {\it et al.}  [UA1 Collaboration],
  %``Search For B0 Anti-B0 Oscillations At The Cern Proton - Anti-Proton
  %Collider. (Paper 2.),''
  Phys.\ Lett.\ B {\bf 186} (1987) 247
  [Erratum-ibid.\  {\bf 197B} (1987) 565].
  %%CITATION = PHLTA,B186,247;%%


\bibitem{Mtop}
  D.~Whiteson  [CDF Collaboration],
hep-ex/0605106.
  %%CITATION = HEP-EX 0605106;%%

\bibitem{hfag}
 Heavy Flavor Averaging Group (HFAG),
  %``Averages of b-hadron properties at the end of 2005,''
hep-ex/0603003 (see page 34).
  %%CITATION = HEP-EX 0603003;%%



\bibitem{cdf}
A.~Abulencia {\it et al.}  [CDF - Run II Collaboration],
  %``Measurement of the B/s0 anti-B/s0 oscillation frequency,''
  Phys.\ Rev.\ Lett.\  {\bf 97} (2006) 062003
  [AIP Conf.\ Proc.\  {\bf 870} (2006) 116]
  [hep-ex/0606027].
  %%CITATION = HEP-EX 0606027;%%




\bibitem{UTA}
  M.~Bona {\it et al.}  [UTfit Collaboration],
  hep-ph/0606167.
  %%CITATION = HEP-PH 0606167;%%
  %%CITATION = HEP-PH 0501199;%%
 J.~Charles {\it et al.}  [CKMfitter Group],
  %``CP violation and the CKM matrix: Assessing the impact of the asymmetric  B
  %factories,''
  Eur.\ Phys.\ J.\ C {\bf 41} (2005) 1
  [hep-ph/0406184].
  %%CITATION = HEP-PH 0406184;%%



\bibitem{onogi}
T.~Onogi,
  PoS {\bf LAT2006} (2006) 017
  [hep-lat/0610115];
  %%CITATION = HEP-LAT 0610115;%% 
M.~Wingate,
  %``B physics on the lattice: Present and future,''
  Mod.\ Phys.\ Lett.\ A {\bf 21} (2006) 1167
  [hep-ph/0604254];
  %%CITATION = HEP-PH 0604254;% S.~Hashimoto,
  Int.\ J.\ Mod.\ Phys.\ A {\bf 20} (2005) 5133
  [hep-ph/0411126];
  %%CITATION = HEP-PH 0411126;%%
 D.~Becirevic, hep-ph/0310072.
  %%CITATION = HEP-PH 0310072;%%




\bibitem{ryan}
N.~Yamada {\it et al.}  [JLQCD Collaboration],
  Nucl.\ Phys.\ Proc.\ Suppl.\  {\bf 106} (2002) 397
  [hep-lat/0110087];
  %%CITATION = HEP-LAT 0110087;%%
A.~S.~Kronfeld and S.~M.~Ryan,
  Phys.\ Lett.\ B {\bf 543} (2002) 59
  [hep-ph/0206058];
  %%CITATION = HEP-PH 0206058;%%  
D.~Becirevic, S.~Fajfer, S.~Prelovsek and J.~Zupan,
  Phys.\ Lett.\ B {\bf 563} (2003) 150
  [hep-ph/0211271].
  %%CITATION = HEP-PH 0211271;%%



\bibitem{casalbuoni}
R.~Casalbuoni  {\it et al.},
%``Phenomenology of heavy meson chiral Lagrangians,''
Phys.\ Rept.\  {\bf 281} (1997) 145
[hep-ph/9605342];
%%CITATION = HEP-PH 9605342;%%
A.~V.~Manohar and M.~B.~Wise,
  %``Heavy quark physics,''
  Camb.\ Monogr.\ Part.\ Phys.\ Nucl.\ Phys.\ Cosmol.\  {\bf 10}, 1 (2000);
  %%CITATION = CMPCE,10,1;%%
J.~F.~Donoghue, E.~Golowich and B.~R.~Holstein,
  %``Dynamics Of The Standard Model,''
  Camb.\ Mon.\ Part.\ Phys.\ Nucl.\ Phys.\ Cosmol.\  {\bf 2} (1992) 1.
  %%CITATION = CMPCE,2,1;%%



\bibitem{seb}
  S.~Descotes-Genon, N.~H.~Fuchs, L.~Girlanda and J.~Stern,
  Eur.\ Phys.\ J.\ C {\bf 34} (2004) 201
  [hep-ph/0311120]; 
  %%CITATION = HEP-PH 0311120;%%
S.~Descotes-Genon, L.~Girlanda and J.~Stern,
  Eur.\ Phys.\ J.\ C {\bf 27} (2003) 115
  [hep-ph/0207337].
  %%CITATION = HEP-PH 0207337;%%



\bibitem{manohar}
  A.~Manohar and H.~Georgi,
%``Chiral Quarks And The Nonrelativistic Quark Model,''
  Nucl.\ Phys.\ B {\bf 234}, 189 (1984).
  %%CITATION = NUPHA,B234,189;%%



\bibitem{pich}
  A.~Pich,
  Rept.\ Prog.\ Phys.\  {\bf 58} (1995) 563
  [hep-ph/9502366].
  %%CITATION = HEP-PH 9502366;%%

\bibitem{gasser}
  J.~Gasser and H.~Leutwyler,
  Nucl.\ Phys.\ B {\bf 250} (1985) 465.
  %%CITATION = NUPHA,B250,465;%%

\bibitem{Ds}
B.~Aubert {\it et al.}  [BABAR],
%``Observation of a narrow meson decaying to D/s+ pi0 at a mass of
%2.32-GeV/c**2,''
Phys.\ Rev.\ Lett.\  {\bf 90} (2003) 242001
[hep-ex/0304021];
%%CITATION = HEP-EX 0304021;%%
D.~Besson {\it et al.}  [CLEO],
%``Observation of a narrow resonance of mass 2.46-GeV/c**2 decaying to  D/s*+
%pi0 and confirmation of the D/sJ*(2317) state,''
Phys.\ Rev.\ D {\bf 68} (2003) 032002
[hep-ex/0305100];
%%CITATION = HEP-EX 0305100;%%
K.~Abe {\it et al.},
%``Measurements of the D/sJ resonance properties,''
Phys.\ Rev.\ Lett.\  {\bf 92} (2004) 012002
[hep-ex/0307052];
%%CITATION = HEP-EX 0307052;%%
E.~W.~Vaandering  [FOCUS Collaboration],
%``Charmed hadron spectroscopy from FOCUS,''
hep-ex/0406044.
%%CITATION = HEP-EX 0406044;%%



\bibitem{Dq}
K.~Abe {\it et al.}  [Belle Collaboration],
Phys.\ Rev.\ D {\bf 69} (2004) 112002
[hep-ex/0307021];
%%CITATION = HEP-EX 0307021;%%
J.~M.~Link {\it et al.}  [FOCUS],
Phys.\ Lett.\ B {\bf 586} (2004) 11
[hep-ex/0312060].
%%CITATION = HEP-EX 0312060;%%



\bibitem{dss}
P.~Colangelo, F.~De Fazio and R.~Ferrandes,
  Mod.\ Phys.\ Lett.\ A {\bf 19} (2004) 2083
  [hep-ph/0407137];
  %%CITATION = HEP-PH 0407137;%%
E.~S.~Swanson,
  Phys.\ Rept.\  {\bf 429} (2006) 243
  [hep-ph/0601110];
  %%CITATION = HEP-PH 0601110;%%  
S.~Godfrey,
  Phys.\ Rev.\ D {\bf 72} (2005) 054029
  [hep-ph/0508078];
  %%CITATION = HEP-PH 0508078;%%
D.~Becirevic, S.~Fajfer and S.~Prelovsek,
  Phys.\ Lett.\ B {\bf 599} (2004) 55
  [hep-ph/0406296].
  %%CITATION = HEP-PH 0406296;%%



\bibitem{norikazu}
  H.~W.~Lin, S.~Ohta, A.~Soni and N.~Yamada,
hep-lat/0607035.
  %%CITATION = HEP-LAT 0607035;%%



\bibitem{green}
  A.~M.~Green {\it et al.}  [UKQCD Collaboration],
  Phys.\ Rev.\ D {\bf 69} (2004) 094505
  [hep-lat/0312007].
  %%CITATION = HEP-LAT 0312007;%%


\bibitem{SUSY-B}
  D.~Becirevic {\it et al.},
  %``B/d anti-B/d mixing and the B/d --> J/psi K(S) asymmetry in general  SUSY
  %models,''
  Nucl.\ Phys.\ B {\bf 634} (2002) 105
  [hep-ph/0112303];
  %%CITATION = HEP-PH 0112303;%%
P.~Ball, S.~Khalil and E.~Kou,
  %``B/s0-anti-B/s0 mixing and the B/s --> J/psi Phi asymmetry in
  %supersymmetric models,''
  Phys.\ Rev.\ D {\bf 69} (2004) 115011
  [hep-ph/0311361];
  %%CITATION = HEP-PH 0311361;%%
A.~J.~Buras {\it et al.}, 
  Nucl.\ Phys.\ B {\bf 659} (2003) 3
  [hep-ph/0210145];
  %%CITATION = HEP-PH 0210145;%%
V.~Barger {\it et al.}, 
Phys.\ Lett.\ B {\bf 596} (2004) 229
  [hep-ph/0405108];
  %%CITATION = HEP-PH 0405108;%%
P.~Ball and R.~Fleischer,
  %``Probing new physics through B mixing: Status, benchmarks and prospects,''
  Eur.\ Phys.\ J.\ C {\bf 48} (2006) 413
  [hep-ph/0604249].
  %%CITATION = HEP-PH 0604249;%%

\bibitem{gabbiani}
  F.~Gabbiani, E.~Gabrielli, A.~Masiero and L.~Silvestrini,
  Nucl.\ Phys.\ B {\bf 477} (1996) 321
  [hep-ph/9604387].
  %%CITATION = HEP-PH 9604387;%%


\bibitem{4f}
R.~Sommer,
hep-lat/0611020; 
  %%CITATION = HEP-LAT 0611020;%%
P.~Dimopoulos {\it et al.}  [ALPHA Collaboration],
hep-lat/0610077.
  %%CITATION = HEP-LAT 0610077;%%
D.~Becirevic {\it et al.}, 
  JHEP {\bf 0408} (2004) 022
  [hep-lat/0401033].
  %%CITATION = HEP-LAT 0401033;%%

\bibitem{giovanni}
  D.~Becirevic and G.~Villadoro,
  Phys.\ Rev.\ D {\bf 70} (2004) 094036
  [hep-lat/0408029].
  %%CITATION = HEP-LAT 0408029;%%

\bibitem{detmold-lin}
  W.~Detmold and C.~J.~Lin, hep-lat/0612028.
  %%CITATION = HEP-LAT 0612028;%%
  
  
\bibitem{sharpe}
S.~R.~Sharpe, hep-lat/0607016;
  %%CITATION = HEP-LAT 0607016;%%
C.~Aubin,
hep-lat/0612013;
  %%CITATION = HEP-LAT 0612013;%%
C.~Bernard {\it et al.},
hep-lat/0611024.
  %%CITATION = HEP-LAT 0611024;%%


\bibitem{QUENCH}
  D.~Becirevic, S.~Prelovsek and J.~Zupan,
  Phys.\ Rev.\ D {\bf 67} (2003) 054010
  [hep-lat/0210048].
  %%CITATION = HEP-LAT 0210048;%%
 %%CITATION = HEP-LAT 0305001;%%


\bibitem{grinstein}
  B.~Grinstein, E.~Jenkins, A.~V.~Manohar, M.~J.~Savage and M.~B.~Wise,
  Nucl.\ Phys.\ B {\bf 380} (1992) 369
  [hep-ph/9204207].
  %%CITATION = HEP-PH 9204207;%%

\bibitem{eeg}
  A.~Hiorth and J.~O.~Eeg,
  Eur.\ Phys.\ J.\ directC {\bf 30} (2003) 006
  [Eur.\ Phys.\ J.\ C {\bf 32S1} (2004) 69]
  [hep-ph/0301118].
  %%CITATION = HEP-PH 0301118;%%
  
\bibitem{dlin}
  D.~Arndt and C.~J.~D.~Lin,
  Phys.\ Rev.\ D {\bf 70} (2004) 014503
  [hep-lat/0403012].
  %%CITATION = HEP-LAT 0403012;%%

\bibitem{springer}
  T.~Mehen and R.P.~Springer,
  Phys.\ Rev.\ D {\bf 72} (2004) 034006
  [hep-lat/0503134].
  %%CITATION = HEP-LAT 0503134;%%

\bibitem{craig}
  G.~Herdoiza, C.~McNeile and C.~Michael  [UKQCD Collaboration],
  Phys.\ Rev.\ D {\bf 74} (2006) 014510
  [hep-lat/0604001]; 
  %%CITATION = HEP-LAT 0604001;%%
C.~McNeile and C.~Michael  [UKQCD Collaboration],
  %``Searching for chiral logs in the static-light decay constant,''
  JHEP {\bf 0501} (2005) 011
  [hep-lat/0411014].
  %%CITATION = HEP-LAT 0411014;%%

\bibitem{bloss}
P.~Colangelo and F.~De Fazio,
%``QCD interactions of heavy mesons with pions by light-cone sum rules,''
Eur.\ Phys.\ J.\ C {\bf 4} (1998) 503
[hep-ph/9706271];
%%CITATION = HEP-PH 9706271;%%
%%CITATION = HEP-PH 0206237;%%
Y.~B.~Dai {\it et al.},
%``Decay widths of excited heavy mesons from QCD sum rules in the leading  order
%of HQET,''
Phys.\ Rev.\ D {\bf 58} (1998) 094032
[Erratum-ibid.\ D {\bf 59} (1999) 059901]
[hep-ph/9705223];
%%CITATION = HEP-PH 9705223;%%
J.~Lu, X.~L.~Chen, W.~Z.~Deng and S.~L.~Zhu,
  %``Pionic decays of D/sj(2317), D/sj(2460) and B/sj(5718), B/sj(5765),''
  Phys.\ Rev.\ D {\bf 73} (2006) 054012
  [hep-ph/0602167];
  %%CITATION = HEP-PH 0602167;%%
  D.~Becirevic, B.~Blossier, Ph.~Boucaud, J.~P.~Leroy, A.~LeYaouanc and O.~Pene,
  %``Pionic couplings g-hat and g-tilde in the static heavy quark limit,''
  PoS {\bf LAT2005} (2006) 212
  [hep-lat/0510017].
  %%CITATION = HEP-LAT 0510017;%%
 %%CITATION = HEP-LAT 0310050;%%

\bibitem{jernej}
  S.~Fajfer and J.~Kamenik,
  Phys.\ Rev.\ D {\bf 74} (2006) 074023
  [hep-ph/0606278].
  %%CITATION = HEP-PH 0606278;%%

}


\end{thebibliography}
\end{document}